# Towards a Formalization of the OSGi Component Framework

Jan Olaf Blech

fortiss GmbH, Munich, Germany

We present a formalization of the OSGi component framework. Our formalization is intended to be used as a basis for describing behavior of OSGi based systems. Furthermore, we describe specification formalisms for describing properties of OSGi based systems. One application is its use for behavioral types. Potential uses comprise the derivation of runtime monitors, checking compatibility of component composition, discovering components using brokerage services and checking the compatibility of implementation artifacts towards a specification.

## 1 Introduction

In this report, we describe a formal definition of syntax and semantics for the OSGi [15] framework. OSGi is a component and service platform for Java. One example of an OSGi system is Eclipse. Eclipse is realized using OSGi with plug-ins as OSGi components. Furthermore, we motivate formalisms for describing properties of OSGi systems.

The idea behind this work is a formalism that is suitable for behavioral types for OSGi. Behavioral types are abstract component descriptions that can be used for checking compatibility, adapting, discovering and checking properties of components at development and at run-time of a system.

Algorithms for comparing specifications based on our semantics as well as a possible tool integration are not in the scope of this paper. A vision paper describing a larger perspective of usages for our work is given in [6].

### 1.1 Overview

We present an overview on OSGi in Section 2. Design choices for a formal component model are presented in Section 3. Our semantics for OSGi and a formal representation is presented in Section 4. Section 5 presents examples of our semantics and discusses typical scenarios. A connection to behavioral types is stated in Section 6. Related work is discussed in Section 7 and a conclusion is featured in Section 8

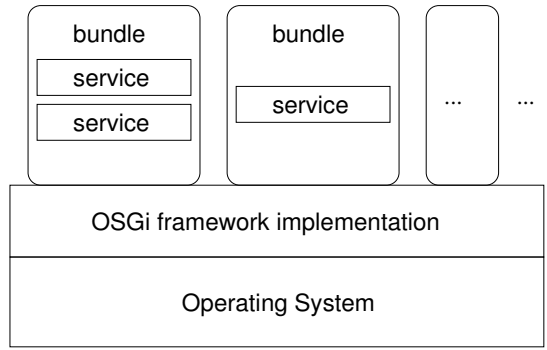

Figure 1: OSGi framework

## 2 Overview on OSGi

We present an overview on OSGi following the description in [6]. The OSGi framework is a component and service platform for Java. It allows the aggregation of services into bundles (cf. Figure 1) and provides means for dynamically configuring services, their dependencies and usages. It is used as the basis for Eclipse plug-ins but also for embedded applications including solutions for the automotive domain, home automation and industrial automation. Bundles can be installed and uninstalled during the runtime. For example, they can be exchanged by newer versions. Hence, possible interactions between bundles can in general not be determined statically.

Bundles are deployed as .jar files containing extra OSGi information. Bundles generally contain a class implementing an OSGi interface that contains code for managing the bundle, e.g., code that is executed upon activation and stopping of the bundle. Upon activation, a bundle can register its services to the OSGi framework and make it available for use by other bundles. Services are technically implemented in Java. The bundle may itself start to use existing services. Services can be found using dictionary-like mechanisms provided by the OSGi framework. Typically one can search for a component with a specified Java interface.

The OSGi standard only specifies the framework including the syntactical format specifying what bundles should contain. Different implementations exist for different application domains like Equinox[1] for Eclipse, Apache Felix[2] or Knopflerfish[3]. If bundles do not depend on implementation specific features OSGi bundles can run on different implementations of the OSGi framework.

---

[1] http://www.eclipse.org/equinox/
[2] http://felix.apache.org/site/index.html
[3] http://www.knopflerfish.org/

# 3 Design Choices for a Formalization of the OSGi Component Model

We describe characteristics of OSGi systems and aspects that need to be covered in a formalization of OSGi and its semantics. Furthermore, we describe aspects that one needs to take care of when describing properties.

**Characteristics of OSGi systems**  Characteristics of OSGi systems comprise the following aspects that we aim to cover in our work:

- Concurrent execution of different methods.
- Blocking method calls.
- Dynamic adding and removing of bundles.
- No global synchronization present in the system in general.

Important for us is the distinction between the OSGi system and its semantics:

- Unlike traditional systems, the OSGi system can change during the execution time as bundles get added and removed.
- The semantics need to handle this changing of the underlying system. Behavior formalized, e.g., as state transitions may become possible or impossible as new bundles are added and removed.

Our formal description of OSGi shall be used for the following purposes (cf. [9, 6]):

- Behavioral types (cf. our work [6]) and
- Runtime verification (cf. our work [5]).

For these applications, we are interested in the operational behavior of OSGi systems.

**Choices for describing properties**  Different design choices for describing properties for OSGi components can be imagined. They may comprise:

- *System and component invariants* may be used to describe properties that shall hold throughout the entire life-time of a system or component.
- *Automata or regular expression like formalisms* may be used to describe sets of possible execution traces. These can be used to define the order of execution events, e.g., distinct method calls. Protocols for component interactions can be defined using these mechanisms.
- *Assertion based formalisms and pre- and post-conditions* may be used to cover the effect of a method call.

# 4 A Formal Component Model

An OSGi component system comprises bundles and objects with their methods as well as a framework to allow communication, e.g., method calls between objects and bundles. The framework provides basic services.

## 4.1 A Method-Call Semantics

In the following we give definitions for formalizing OSGi. We concentrate on a semantics that captures behavior originating from method calls between different bundles. Memory and exchange of data between these bundles and objects is not yet taken into account. Thus, we provide an overapproximation and abstraction of a real system.

**Definition 1 (Object and method definitions)** *An object is a tuple* $(m_0, ..., m_n)$ *comprising method definitions* $m_0, ..., m_n$.

The semantics of an object is given by the semantic interpretation of its methods and its object state.

**Definition 2 (Semantics of a Method)** *The semantics of a method is giving by an automaton* $(L, E, l_0)$ *comprising a set of locations $L$ an initial location $l_0 \in L$ and edges $E = (l_i, M, l_j)$ between locations.*

An addition to source and target location $l_i$ and $l_j$ an edge comprises a set (can be ordered) of method calls and special calls $M$. These can be tuples $(m, o, b) \in M$ comprising a method definition $m$ of an object $o$ that is associated with a bundle $b$. Furthermore, $M$ can contain special calls for:

- Adding and removing bundles.
- Creating and deleting objects.

Each transition from $E$ represents an action that is atomic or non-terminating to the method but not to the OSGi system. It can represent a memory update, but also other method calls. A method call can itself trigger a non-terminating method in the same or in other objects, so transitions do not have to terminate.

**Definition 3 (Object state and method status states)** *An object state is a set of tuples*

$$s_o = \{(m_n, l_{n_i}, id_n, cs_n), ..., (m_p, l_{p_j}, id_p, cs_p)\}$$

*comprising active method status states representing method calls: method definitions, their actual locations, an $id$ and a call state $cs$*

$$(m_n, l_{n_i}, id_n, cs_n), ..., (m_p, l_{p_j}, id_p, cs_p).$$

The call state is part of an active method status state. It is a set of method definitions and method id plus status information for which the active method is waiting to return.

The id is used to distinguish different calls to the same method.

Bundles are aggregate objects into units that are enumerated in the OSGi system and can be loaded and removed during runtime by user commands or from other bundles.

**Definition 4 (Bundle)** *A bundle is a set of objects $\{o_{activator}, ...\}$ comprising an object $o_{activator} = (m_{start}, m_{stop}, ...)$ which is created on activation. It comprises two distinct methods $m_{start}, m_{stop}$ which are called during activation and deactivation.*

In an implemented OSGi system, the $o_{activator}$ object has to implement the `BundleActivator` interface defined in `org.osgi.framework`. It comprises two methods with signatures:

```
void start(BundleContext context) throws java.lang.Exception
```

and

```
 void stop(BundleContext context) throws java.lang.Exception
```

The semantical definition of bundle states aggregates its object states.

**Definition 5 (Bundle state)** *A bundle state is defined as a set of object states $s_b = \{s_{o_i}, ..., s_{o_k}\}$ for object states $s_{o_i}, ..., s_{o_k}$.*

A standard OSGi system has one (Equinox) or more bundles which are active at startup.

**Definition 6 (OSGi system)** *An OSGi system is a set of bundles. It comprises a distinct bundle $b_{init}$ which is activated at start-up.*

Analog to object and bundle state, we define an OSGi system state.

**Definition 7 (OSGi system state)** *A system state is defined as a set of bundle states $s = \{s_{b_i}, ..., s_{b_k}\}$ for bundle states $s_{b_i}, ..., s_{b_k}$.*

The initial state of an OSGi system comprises the start of the $start$ method in the activator object of the initial bundle.

**Definition 8 (Initial state)** *The initial state of an OSGi system is defined as $s_{init} = \{s_{b_{init}}\}$ with $s_{b_{init}} = \{o_{activator}\}$ and $o_{activator} = \{(m_{start}, l_{start_0}, 0, \emptyset)\}$.*

## 4.2 Dynamic Architecture of OSGi System

An important aspect of our formalization is the impact on OSGi operations that can change the structure of OSGi systems. Such operations can be triggered by OSGi methods themself (cf. Definition 2), e.g., comprising adding and removing objects and bundles. Another option is to perform these operations by a command line interface (e.g., starting eclipse with the console option using Equinox) at runtime on the OSGi framework.

Here, We distinguish the following structure changing operations on OSGi systems:

- Starting / Loading a system.
- Adding a bundle and activating it.
- Removing a bundle (and deactivating it).
- Adapting a bundle and its services.
- Closing / Removing a system.

Characteristic for these operations is the fact that new behavior becomes possible or is removed at runtime of the OSGi system. Thus, the semantics of an OSGi system and possible events can in general not be determined statically at start up of a system.

### 4.3 State Transitions in OSGi

Here, we present our state transition definition. State transitions can modify both, structure of a system and the classical state.

State transitions are made up from local transitions appearing within methods and from handling terminated methods. In general state transitions are highly non-deterministic and define a relation of

$$\text{previous system state} \times \text{previous system definition} \times$$
$$\text{next system state} \times \text{next system definition}$$

**Definition 9 (Global State Transition)** *For an OSGi system $S = \{..., b, ...\}$: Given the system state $s = \{..., s_b, ...\}$ with $s_o \in s_b$ and $(m, l_i, id, cs) \in s_o$ the following basic state transition cases can be distinguished:*

- *Calling a method $m'$ of object $o'$ from bundle $b'$: We regard a transition $(l_i, M, l_j) \in o$ with $o \in b$. The following steps are performed.*
  1. *The step can be performed under the preconditions that $(m', o', b') \in M$ and $o'$ and $b'$ exist in $S$.*
  2. *$cs$ is updated by adding the method call indicating its bundle, object and id.*
  3. *A new element $(m', l'_0, id', \emptyset)$ is added to the object state where $m'$ belongs to. $id'$ is a new identifier for the method $m'$.*

- *Executing a method step: We regard a transition $(l_i, M, l_j) \in o$ with $o \in b$.*
  1. *The step can be performed under the precondition that $cs = \emptyset$.*
  2. *$s_o$ is updated as $s'_o = s_o / (m, l_i, id, cs) \cup \{(m, l_j, id, cshandle(M))\}$. Thus, $(m, l_i, id, cs)$ is removed and $(m, l_j, id, cshandle(M))$ is added instead.*
  
    *$cshandle$ transforms $M$ into a representation that indicates which methods have been called and keeps track of its id. Furthermore, $cshandle$ takes care of special operations that modify the system definition.*

- *Returning from a method call: Any method status state with $cs = \emptyset$ and no edge that my lead to a possible succeeding state can be processed in the following way:*
    1. The method status state is removed.
    2. The call state of any method that $m$ has called is updated such that the entry for the $m$ call is removed.

*Furthermore, the following operations are handled:*

- *Adding a bundle : The $cs$ from any object state $s_o$ with $(m, l_i, id, cs) \in s_o$ can contain a special operation (denoted: add bundle $b'$) for adding a bundle $b'$ and changing the system definition from $S$ into $S' = S \cup \{b'\}$.*

- *Removing a bundle: The $cs$ from any object state $s_o$ with $(m, l_i, id, cs) \in s_o$ can contain a special operation for removing a bundle $b'$ (denoted: remove bundle $b'$) and changing the system definition from $S$ into $S' = S/\{b'\}$.*

- *Creating an object. The $cs$ from any object state $s_o$ with $(m, l_i, id, cs) \in s_o$ can contain a special operation (denoted: create object $(o', b)$) for adding an object $o'$ and changing a bundle definition $b \in S$ to $b' = b \cup \{o'\}$. The system definition is, thus, changed from $S$ into $S' = S/b \cup \{b'\}$.*

- *Deleting an object: The $cs$ from any object state $s_o$ with $(m, l_i, id, cs) \in s_o$ can contain a special operation (denoted: delete object $(o', b)$) for deleting an object $o'$ and changing a bundle definition $b \in S$ to $b' = b/o'$. The system definition is, thus, changed from $S$ into $S' = S/b \cup \{b'\}$.*

Note, that as an additional constraint objects may only be created in the same bundle as the object of the method that is triggering the creation belongs to.

## 5 Examples for OSGi System Formalizations

In this section we present examples of our OSGi formalization thereby giving some evidence of the usability of our semantics and system definition.

### 5.1 Example: On Startup

Figure 2 shows a typical start method of an activator object (cf. Definition 4) in a bundle. The method is defined using Definition 2 as:

$(\{l_0, l_1, l_2, l_3\},$
  $\{(l_0, \{createobject(service1, servicebundle)\}, l_1),$
  $(l_1, \{(initialize, service1, servicebundle)\}, l_2),$
  $(l_2, \{deleteobject(service1, servicebundle)\}, l_3)\},$
 $l_0)$

The startup method creates an object $service1$ in a bundle. It calls an initialization method and deletes the object. In our semantics we have no timing constraints. Thus, after the creation of the object and after the initialization call, the execution of the startup method may be delayed infinitely, but may also continue directly. This kind of semantics is similar to the BIP [16] semantics and is suitable for defining and reasoning about safety properties based on the synchronization of concurrent automata at distinct points in execution.

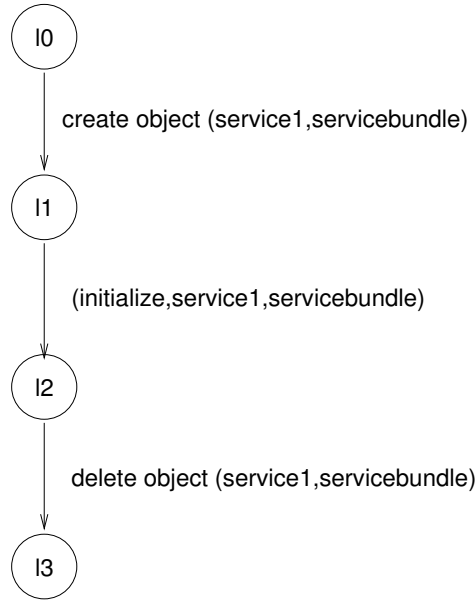

Figure 2: A typical start method

An example specification of the initialization method is shown in Figure 3. Note, that the call of the initialize method establishes a synchronization between the two automata specifying the startup and the initialize method: after the transition it is guaranteed that both are in the locations $l_0$ and $l_3$ respectively. In the presents of other methods / objects / bundles, this synchronization is lost directly after the call, since our state transition relation (Definition 9) allows arbitrary transitions within the OSGi system as next state transitions.

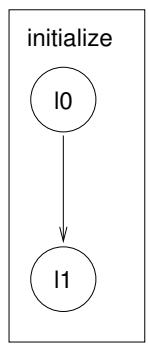

Figure 3: A simple initialization

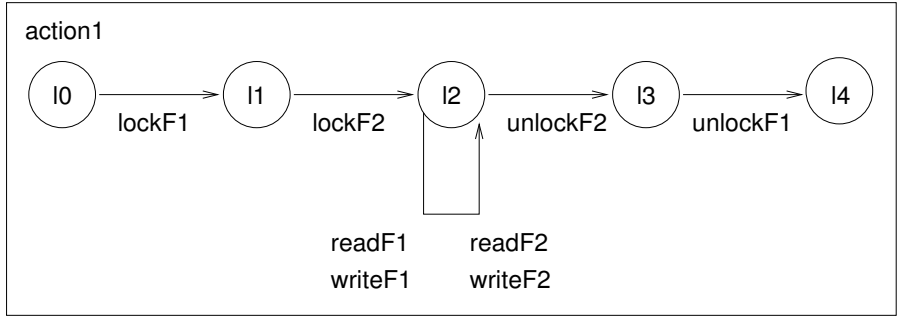

Figure 4: Service Action 1

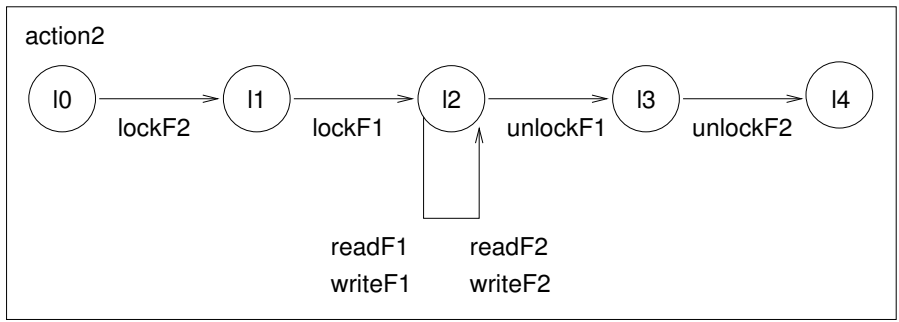

Figure 5: Service Action 2

## 5.2 Example: Objects Interacting

Let's assume that the created object offers two methods *action1* (Automaton definition in Figure 4) and *action2* (Automaton definition in Figure 5). In an implemented system both actions make use of two resources: files. These are represented by other OSGi bundles or objects. Here, the automata specify methods that are called to lock and unlock files by the users of the files. Furthermore, possible operations in them – read and write – are used. It can be seen that the two action methods lock the files in a different order. In an implemented system deadlocks can be possible if the two actions are executed in parallel and both have acquired one lock each. The deadlock possibility can not be determined by the specification of the methods alone, since we only talk about outgoing calls and not about required protocols enforced, e.g., by an objects memory state and guard conditions on transitions. Thus, the need for additional specification elements arises here.

The method of an object for handling files are shown in Figure 6. The method definitions are rather simple since only distinct operations on memory are performed and we do not cover these in the introduced semantics.

The required protocol of using such a file object can not be seen from Figure 6. We have, however, depicted it in Figure 7: It is shown that a file needs to be put in a lock state before operations can be carried out. Then it can be unlocked again. In addition to what is shown in the figure one can require that only one other method can hold a lock at the same time. Furthermore, complementing our OSGi specifications which talk about outgoing method calls, this additional constraints describes incoming method calls. Specifications on incoming and outgoing method calls, together, define a protocol which one can formally handle and, e.g., check for potential deadlocks.

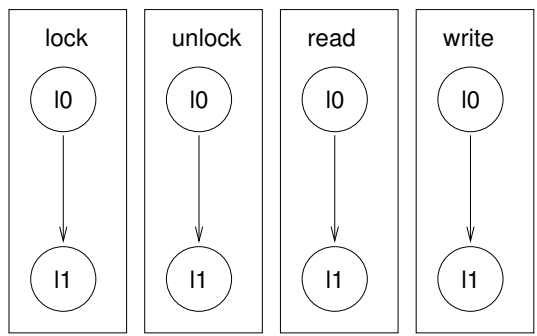

Figure 6: Method semantics of file operations

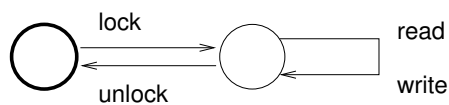

Figure 7: Intended interplay of method calls

**Some semantic artifacts and their representation** Our semantics provides an abstract view on OSGi systems. Some formalization issues that are not discussed above are explained in the following:

- *Recursive method calls.* These are treated by allowing different calls to the same method. Several versions of the methods execution states in an object state can be realized with means to distinguish them: We enumerate them using different ids.

- *Calling the same method from different objects concurrently.* This case is treated in a similar way: different ids are used for different method calls.

- *Methods mutually calling each other.* Different instances of methods are created. Again they are distinguished by a new id each time a method is called.

- *Objects and classes.* Our semantics is defined upon objects. The concept of classes is orthogonal to the presented definitions and can be introduced. Thereby additional constraints can be put on creating objects and calling their methods. Thus, the semantics becomes more fine-grained.

- *Exceptions.* Exceptions are represented by adding transitions to the automata that define a method's semantics. These transitions represent the extra control flow that is induced by a thrown exception. Transitions contain the creation of an object that represent the exception.

## 5.3 Semantics for Memory Updates and Guards

The semantics definition from Section 4.1 takes only method calls and their order into account. States can be augmented with memory and their updates. Transitions can be augmented with guards based on memory. This provides a more fine-grained semantics. The semantics presented in Section 4.1 is a strict overapproximation, thus, has in general more non-determinism than a semantics that takes memory into account.

In order to add memory support, the following changes are needed:

- Memory needs to be added object states.

- Guards that are defined upon memory need to be added to transitions.

- Updates defined as mappings from previous memory state to succeeding memory state need to be added to transitions.

- A function that performs exchange of method parameters (also between different bundles) needs to be added to a method definition. It is used during the method call case.

- Analog: A second function that performs exchange of return values must be added to a method definition. It is used in the method return case.

**Additional aspects**  Some aspects are not properly covered by the described semantics. These comprise:

- *Reflection within Java.* Handling reflection in our semantics means that a method can call methods of objects without knowing their names and parameters at compile time. This can be realized by adapting the semantics definition of methods at runtime, i.e., adding and removing edges and locations.

- *Multiple threads within a single method call.* This could be realized by introducing parallel automata as definitions for the semantics of methods.

Common to these aspects is the fact that their handling in real OSGi implementations can be implementation dependent.

## 5.4 Interpretation of the Semantics

Here, we highlight some properties of our semantics and system definitions. The definitions from Section 4 define the semantics of an OSGi system as a state transition system. Building on this, we can think of derived formalisms like traces and coinductive structures. Relating semantics can be done using (weak-)simulation and bisimulation based definitions.

**Non-determinism**  In Section 5 it is motivated that systems are usually specified with a very high degree of non-determinism. If two methods run in parallel it is possible that one does an arbitrary finite number of transitions from locations to other locations before the other one performs a single transition. OSGi in its standard form imposes the same behaviors: it does not offer guarantees regarding timing-constraints and the order of execution of system parts that are not communicating. There is, however, work on extensions for real-time applications of OSGi using real-time Java (e.g., [2]). Thus, our formalization captures this degree of non-determinism introduced by the absence of any timing-guarantees.

**Synchronization points**  Methods run independently without synchronization if they do not call other methods. Each method call is blocking, thus a method waits until the called method is finished. This implies a synchronization between the called and the calling methods. The synchronization is only between these methods. If memory is involved between the two objects the methods belong to, there is synchronization between these objects, too. Other objects and methods need not be effected (although additional method calls involving other objects may be triggered). Thus, in general there is no global synchronization, between the components of an OSGi system.

**Outgoing calls**  Our semantics captures the order of outgoing method calls within an OSGi system. Constraints on the order of incoming method calls or changes to a memory state are not captured. Furthermore, timing dependencies between method calls are not taken into account. Modeling timing features would require the definition of a timing component. This can be modeled by defining that a periodic method call corresponds to the evolving of a distinct time period. We have described similar modeling techniques in [7] and[4].

# 6 Abstractions and Behavioral Types

Section 4.1 and Section 5.3 describe and suggest two levels of abstraction for a semantic view on OSGi: state transitions systems taking method calls into account as well as an extension that regards memory. Additional levels are possible. For behavioral types we are usually interested in more abstract views. On the other hand for a more implementation centric view more concrete views than the semantics at hand might be needed. Here, we sketch some specification formalisms that capture aspects of OSGi and can be based on the semantics described above.

## 6.1 Protocol Specification

One purpose of our work is the description of interaction protocols between different entities like OSGi bundles. Here, we propose to describe interaction protocols between bundles, objects and methods as automata (cf. e.g., interface automata [1]) or regular expressions.

**Interaction protocols for bundles and objects**  Objects and bundles can register a service protocol – describing, e.g., incoming method calls – that they expect. This can be done by using:

- *Regular expressions.* Thereby bundles and objects can indicate expected events. Events can be incoming or outgoing method calls. Thus, the regular expression specifies their order. Regular expressions are terms over an alphabet of events using the $+$ for alternatives, the . for concatenation and the $*$ as the star operator.

- *Finite automata.* Regular expression can be described by an equivalent finite automaton, too. We define our finite automata as a set of locations, an initial location and a transition relation comprising a predecessor and a successor location labeled with an event.

While in our applications the event is typically a method call or a set of method calls, other possibilities like timing events, or creation and deletion of objects and bundles are also possible.

For example the protocol given in Figure 7 can be described as a regular expression as follows:

$$((\text{INC: Lock}).(\text{INC: Read} + \text{INC: Write})^*.(\text{INC: Unlock}))^*$$

The expression describes a sequence that can be repeated. It starts with a lock and ends with an unlock. Between lock and unlock an arbitrary number of read and write operations can occur. The INC denotes expected incoming method call.

The actions from Figure 6 and Figure 7 describe outgoing method calls. This can be written using our regular expressions as:

$$(\text{OUT: LockF1}).(\text{OUT: LockF2}).$$
$$((\text{OUT: ReadF1}) + (\text{OUT: ReadF2}) + (\text{OUT: WriteF1}) + (\text{OUT: WriteF2}))^*.$$
$$(\text{OUT: UnlockF2}).(\text{OUT: UnlockF1})$$

and

$$(\text{OUT: LockF2}).(\text{OUT: LockF1}).$$
$$((\text{OUT: ReadF1}) + (\text{OUT: ReadF2}) + (\text{OUT: WriteF1}) + (\text{OUT: WriteF2}))^*.$$
$$(\text{OUT: UnlockF1}).(\text{OUT: UnlockF2})$$

One can now use these protocol specifications, e.g., for checking:

- *Compatibility* This addresses the question if the operations that one object expects to be called are called by another object. Furthermore, the correct order of calls is of interest.

- *Additional properties* Properties that relate distinct semantical aspects of bundles and objects are of interest. In the given example, the question arises whether a deadlock can occur or not.

In order to perform these checks and analysis one has to match elements of a specification for one component with elements of a specification from another component. In the given example the protocol comparison has to deal with two instances of a file component and has – for example – to relate the (OUT: LockF1) and (OUT: LockF2) with instances of (INC: Lock).

**Parameterized specifications** For facilitating the relation of specifications we define parameterized specifications. These comprise:

- *Parameterized regular expressions.* Here, each event used in a regular expression can be augmented with a parameter. For our example file component specification this results in the following expression, parameterized with $<F>$.

  $((\text{INC: Lock}<F>).(\text{INC: Read}<F> + \text{INC: Write}<F>)^*.(\text{INC: Unlock}<F>))^*$

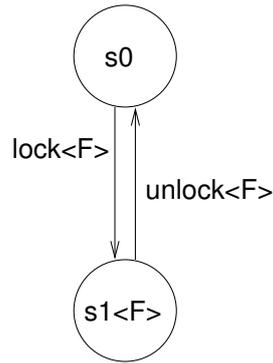

Figure 8: Parameterized specification for locking / unlocking

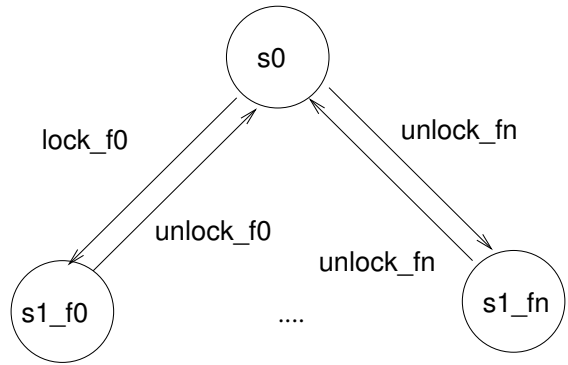

Figure 9: Instantiating a parameterized specification

- *Parameterized automata.* Similar to regular expressions, locations and events in transitions of automata can be augmented with parameters.

Instantiation is done, by substituting concrete values for the parameter. Instantiation of parameters is dependent on concrete application scenarios.

**Example instantiations of parameterized specifications** We regard two kinds of instantiations as particularly useful for describing a protocol of expected incoming method calls.

Consider the refined version of Figure 7 in Figure 8 for locking and unlocking a resource. The lock state as well as the method calls that lead to the lock state are parameterized.

- A first instantiation is shown in Figure 9. Here, the parameter is instantiated by instances $f0, ..., fn$. Each of them gets its own lock state and its own method call that lead to this lock state.

- In case only one lock state is wanted, one can still deal with different parameterized method calls and use the instantiation shown in Figure 10

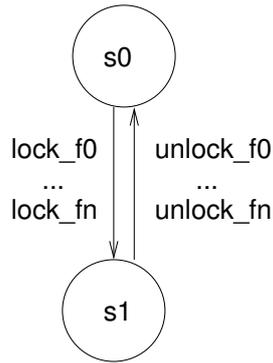

Figure 10: Another way of instantiating a parameterized specification

## 6.2 Invariants of Bundles and OSGi Systems

Component invariants can be used to describe objects, bundles and systems.

**Guaranteeing system invariants** A system invariant can be stated. Each adding and removing of components must ensure that the invariant is preserved.

An invariant is a predicate defined on the state of an OSGi system and the system itself. It is defined using a specification mechanism. Here, we suggest to base specification mechanisms on propositional logic: conjunctions, disjunctions, negations, implications and atoms. Atoms are predicates themself. They capture properties about systems and system states and can contain the following operations as ingredients:

- Checks for bundle and object inclusion in a system.
- Checks for object state and method inclusion in a system state.
- Checks for active method locations in a system state.
- In the case of memory: checks and relation of data in the memory.

These ingredients are combined and can form large invariants that can be local, i.e., the artifacts in the atoms check only for properties of an object or a bundle or global such that facts on several bundles or objects or the entire system is used in the atoms.

This suggested invariant specification mechanism is similar to the D-Finder [3] invariants for the BIP system [16].

## 7 Related Work

The described semantics bears similarity to the semantics of the BIP framework [16]. The idea that different parts of a system can be described using concurrent automata and synchronization between components happen during distinct state transitions is featured in BIP. These synchronizations are realized during method calls in our semantic definition. An extension of BIP for modeling dynamic architectures is described in [10]. A formalization suitable for a higher-order theorem prover of these aspects is described in [8].

To our knowledge existing work does describe OSGi and its semantics only at a very high level. A specification based on process algebras is featured in [17]. Some investigations on the relation between OSGi and some more formal component models have been done in [14]. Aspects on formal security models for OSGi have been studied in [12]. JML [11] provides assertions, pre- and postconditions for Java programs. It can be used to specify aspects of behavior for Java methods.

We have advocated behavioral types in [9] and described a possible framework for OSGi and runtime verification in [6]. Interface automata [1] are one form of behavioral types that is similar to the one our component model is to be used with. Behavioral types have also been used in the Ptolemy framework [13].

## 8 Conclusion

We presented a formalization of OSGi and its semantics. Based on this, we presented formalisms for describing properties of OSGi systems. The semantics and the properties are intended to be used together with behavioral types [6, 9] and related work [5] we are conducting.